\documentclass[12pt]{article}

\title{Mark-Recapture with Multiple Non-invasive Marks}

\author{Simon Bonner and Jason Holmberg}

\date{\today}

\usepackage{setspace,amsmath,natbib,bm,enumerate,longtable,graphicx,rotating,url}
\usepackage[nolists]{endfloat}
\usepackage[margin=1.25in]{geometry}

\usepackage{lineno}
\newcommand*\patchAmsMathEnvironmentForLineno[1]{%
  \expandafter\let\csname old#1\expandafter\endcsname\csname #1\endcsname
  \expandafter\let\csname oldend#1\expandafter\endcsname\csname end#1\endcsname
  \renewenvironment{#1}%
     {\linenomath\csname old#1\endcsname}%
     {\csname oldend#1\endcsname\endlinenomath}}% 
\newcommand*\patchBothAmsMathEnvironmentsForLineno[1]{%
  \patchAmsMathEnvironmentForLineno{#1}%
  \patchAmsMathEnvironmentForLineno{#1*}}%
\AtBeginDocument{%
\patchBothAmsMathEnvironmentsForLineno{equation}%
\patchBothAmsMathEnvironmentsForLineno{align}%
\patchBothAmsMathEnvironmentsForLineno{flalign}%
\patchBothAmsMathEnvironmentsForLineno{alignat}%
\patchBothAmsMathEnvironmentsForLineno{gather}%
\patchBothAmsMathEnvironmentsForLineno{multline}%
}

\newcommand{\T}[1]{{#1}^T}

\newcommand{\bA}{\bm{A}'}
\newcommand{\nullA}{\mbox{null}(\bA)}
\newcommand{\bp}{\bm{p}}
\newcommand{\bmf}{\bm{f}}
\newcommand{\bphi}{\bm{\phi}}
\newcommand{\btheta}{\bm{\theta}}
\newcommand{\bthetacurr}{\btheta^{curr}}
\newcommand{\logit}{\mbox{logit}}
\newcommand{\Var}{\mbox{Var}}
\newcommand{\ntot}{N}
\newcommand{\ntotcurr}{N^{curr}}
\newcommand{\ntotprop}{N^{prop}}
\newcommand{\nobs}{f}
\newcommand{\ntrue}{x}
\newcommand{\ntruecurr}{\ntrue^{curr}}
\newcommand{\ntrueprop}{\ntrue^{prop}}
\newcommand{\bnobs}{\bm \nobs}
\newcommand{\bntrue}{\bm \ntrue}
\newcommand{\bntruecurr}{\bntrue^{curr}}
\newcommand{\bntrueprop}{\bntrue^{prop}}

\newcommand{\mobslink}{L}%m^{obs}}
\newcommand{\mtruelink}{K}%m^{true}}
\newcommand{\mobs}{L^{'}}%m^{obs}}
\newcommand{\mtrue}{K^{'}}%m^{true}}
\newcommand{\lhistk}{l(k)}
\newcommand{\rhistk}{r(k)}

\newenvironment{keywords}%
{\noindent {\bf Keywords:}}
{}

\renewcommand{\paragraph}[1]{}

\begin{document}

% Title page

%% Non-biometrics
\maketitle
\doublespacing

%  This label and the label ``lastpage'' are used by the \pagerange
%  command above to give the page range for the article

\label{firstpage}

\linenumbers

% Abstract
\begin{abstract}
  Non-invasive marks, including pigmentation patterns, acquired
  scars,and genetic markers, are often used to identify individuals in
  mark-recapture experiments. If animals in a population can be
  identified from multiple, non-invasive marks then some individuals
  may be counted twice in the observed data. Analyzing the observed
  histories without accounting for these errors will provide incorrect
  inference about the population dynamics. Previous approaches to this
  problem include modeling data from only one mark and combining
  estimators obtained from each mark separately assuming that they are
  independent. Motivated by the analysis of data from the ECOCEAN
  online whale shark ({\it Rhincodon typus}) catalog, we describe a
  Bayesian method to analyze data from multiple, non-invasive marks
  that is based on the latent-multinomial model of
  \citet{Link2010}. Further to this, we describe a simplification of
  the Markov chain Monte Carlo algorithm of \citet{Link2010} that
  leads to more efficient computation. We present results from the
  analysis of the ECOCEAN whale shark data and from simulation studies
  comparing our method with the previous approaches.
\end{abstract}

\medskip

%%% Local Variables: 
%%% mode: latex
%%% TeX-master: "../multiple_marks_ms_biometrics"
%%% End: 

% \maketitle

% Introduction

\section{Introduction}
\label{sec:introduction}
Non-invasive marks (also called natural marks) include patterns in
pigmentation, genetic markers, acquired scars, or other natural
characertistics that allow researchers to identify individuals in a
population without physical capture. Visible marks have long been used
to identify individuals of some species that are hard to tag,
particularly marine mammals, and non-invasive marks are now being used
more widely. \citet{Yoshizaki2009} and \citet{Yoshizaki2011} reference
studies including:
\begin{itemize}
\item studies based on photographs of large cats (cheetahs, snow
  leopards, and tigers),
\item scar patterns on marine mammals (manatees and whales),
\item skin patterns of reptiles and amphibians (snakes, crocodiles,
  and salamanders),
\item and genetic marks in various species (bears, wombats, and
  whales).
\end{itemize}
The primary advantage of non-invasive marks over man-made marks is
that they can often be observed from a distance or through the
collection of secondary material (e.g.~hair samples or scat). This
means that individuals can be identified passively without physical
contact. Further, many non-invasive marks allow every individual in
the population to be identified from birth. However, mark-recapture
data collected from non-invasive marks can present several modeling
challenges. Previous statistical developments have considered that
non-invasive marks may be misidentified at non-negligible rates
\citep{Lukacs2005,Wright2009,Yoshizaki2011}, that individuals' marks
may change over time \citep{Yoshizaki2009}, and that some non-invasive
marks (e.g scar patterns) may be restricted to a subset of the
population \citep{Da-Silva2003,Da-Silva2006}. We consider the problem
of modeling the demographics of a population from mark-recapture data
when individuals have been identified from multiple, non-invasive
marks.

\paragraph{Non-invasive Marks}
The specific application we consider concerns modeling the aggregation
of whale sharks ({\it Rhincodon typus}) in Ningaloo Marine Park (NMP),
off the west coast of Australia. Whale sharks aggregate at NMP each
year between April and July. During this time, whale sharks are
located by tour companies and photographs are taken by tourists and
tour operators who upload their images to the online ECOCEAN whale
shark library. Whale sharks can be identified by the unique pattern of
spots on their flanks, and computer assisted methods are used to match
photographs in the library. Matches are then used to generate capture
histories which provide information about the timing of the sharks'
arrival and departure from NMP and their survival across years
\citep[see][for further details]{Holmberg2009}.

\paragraph{Multiple Non-invasive Marks}

The challenge in modeling this data is that sharks may be photographed
from either the left or the right side, but the spot patterns are not
the same. This means that photographs from the two sides of a shark
cannot be matched without further information. In particular, the spot
patterns on the right and left can only be matched if the shark was
photographed from both sides during one encounter or more. If this has
not happened then photographs of the same shark taken from different
sides on different occasions cannot be linked and the shark will
contribute two separate histories to the observed data. Ignoring this
problem and naively modeling the observed encounter histories will
inflate the apparent number of sharks identified and create dependence
between the encounter histories. This violates a key assumption of
most mark-recapture models. One solution is to construct encounter
histories based on photographs from either the left or right side
alone, but this removes information from the data. As an alternative,
\citet[][pg.~294]{Wilson1999} suggests combining inferences obtained
from left- and right-side photographs of bottlenose dolphins by
averaging separate point estimates and computing standard errors
assuming that these estimates are independent. The bias of the
combined estimate is the average of the biases of the individual
estimates (the combined estimate is unbiased if the individual
estimates are unbiased), but the assumption of independence is
violated and standard errors will be underestimated. More recently,
\citet{Madon2011} describes a method to estimate abundance from
multiple marks by adjusting the sufficient statistics required to
compute the Jolly-Seber estimator, but we have concerns with this
method. Though the observed counts underestimate some of the
statistics and overestimate others, \citet{Madon2011} uses the same
adjustment factor for all and constrains its value to be between 0 and
1. Simulations \citet{Madon2011} presents indicate a clear problem in
that the coverage of confidence intervals is much lower than their
nominal value, even when the population is large and the capture
probability is close to 1. These issues are discussed further in
\citet{bonner2012}. We are also aware of methods similar to ours being
developed concurrently by \citet{mcclintock2013}.

The primary contribution of our work is to provide a valid method of
modeling a population's dynamics using data from multiple,
non-invasive marks. We do so by constructing an explicit model of the
observation process that allows for multiple marks and applying
Bayesian methods of inference via Markov chain Monte Carlo (MCMC)
sampling. Our model is a modification of the latent multinomial model
(LMM) presented in \citet{Link2010} for modeling mark-recapture data
based on genetic marks with non-negligible misidentification
rates. Further to this, we provide a more efficient simplification of
the MCMC algorithm of \citet{Link2010}.

% Section \ref{sec:data} describes the whale shark data and Section
% \ref{sec:methods} gives details on the structure of our model and our
% methods of inference. We provide results from a simulation study and
% the application of our method to the whale shark data in Sections
% \ref{sec:simulation-study} and \ref{sec:application}, and conclude
% with a discussion of the method and further extensions in Section
% \ref{sec:conclusion}.

%%% Local Variables: 
%%% mode: latex
%%% TeX-master: "../multiple_marks_ms_biometrics"
%%% End: 

% Data

\section{Data}
\label{sec:data}

Data for our analysis were obtained from the ECOCEAN on-line whale
shark library (available at \url{www.whaleshark.org}). This library
contains photographs of whale sharks taken by recreational divers and
tour operators worldwide and submitted electronically. The library has
been operational since 2003, and more than 41,000 photographs had been
submitted by over 3,300 contributors as of January, 2013.

New photographs submitted to the library are matched against existing
photographs using two computer algorithms
\citep{arzoumanian2005,vantienhoven2007}. Identities are based on the
pattern of spots on the flank, believed to be unique, and the
algorithms operate independently using significantly different
approaches to provide complementary coverage in evaluating
matches. All matches generated by the computer algorithms are
confirmed by two or more trained research staff to minimize the
probability of false matches. Further details on the study site, the
observation protocols, and the algorithms for matching photographs are
provided in \citet{Holmberg2009}.

We model only the data collected from the northern ecotourism zone of
NMP during the 16 week period between April 1 and July 31, 2008. This
period was divided into 8 capture occasions of 2 weeks each, and
sharks may have been encountered multiple times during a single
capture occasion. Five possible events may occur; on each occasion, a
shark may:
\begin{enumerate}[1)]
\item not be encountered at all (event $0$)
\item be photographed from the left only (event $L$),
\item be photographed from the right only (event $R$),
\item be photographed from both sides simultaneously on at least on
  encounter (event $S$), or
\item be photographed from both sides but never simultaneously (event
  $B$).
\end{enumerate}
We will denote a generic encounter history made from these events by
$\bm{\omega}$. 

Problems with identification arise because the pattern of spots on the
left and right flanks are not the same. It is only possible to match
the skin patterns from the two sides of a shark if photographs of both
sides were taken simultaneously during at least one capture occasion
-- i.e., there is at least one $S$ in its encounter
history. Otherwise, an individual photographed from both sides will
contribute two encounter histories to the data set -- one containing
the observations of its right side and the other containing the
observations of its left side.

Suppose, for example, that an individual's true encounter history is
00L0B0R0. This history is not observable because the two sides of the
individual were never photographed simultaneously. Hence, the
individual will contribute two observed histories to the data --
00L0L000 and 0000R0R0. Working backward, the observed histories
00L0L000 and 0000R0R0 may either come from one individual encountered
on three occasions or from two separate individuals each encountered
on two two or more occasions.

For a study with $T$ capture occasions there are $5^T-1$ possible true
capture histories (we condition on capture and ignore the zero
history). Of these, $(5^T-1)-(4^T-1) + 2(2^T-1)$ histories can be
observed. These include the $(5^T-1)-(4^T-1)$ that contain at least
one $S$, which we call simultaneous histories, the $2^T-1$ histories
that include only $0$ and $L$, left-only histories, and the $2^T-1$
histories that include only $0$ and $R$, right-only histories. The
remaining $(4^T-1) - 2(2^T-1)$ contain either $L$ and $R$ together
and/or $B$ but no $S$ and cannot be observed. Individuals with these
true histories contribute two observed histories to the data. When a
left-only and right-only history, call them $\bm \omega_L$ and $\bm
\omega_R$, combine to form a third history, $\bm \omega_C$, we say
that $\bm \omega_L$ and $\bm \omega_R$ are the left and right parents
of child $\bm \omega_C$.

% Consider, for example, the observed histories 00L0L000 and
% 000RR000. These histories may come from two separate individuals, one
% that was photographed from the left only on two occasions and one that
% was photographed from the right only on two occasions, or they may
% come from a single individual with history 00LRB000. There are $5^{8}
% > 390,000$ possible capture histories from a study with 8
% occasions. Of these, 65,026 (16.6\%) cannot be observed because they
% contain observations from both the left and right sides without
% simultaneous photographs, and any individual with one of these
% histories will contribute two observed histories to the data set.

%%% Local Variables: 
%%% mode: latex
%%% TeX-master: "../multiple_marks_ms_biometrics"
%%% End: 

% Methods

\section{Methods}
\label{sec:methods}

\subsection{Latent Multinomial Model}

% With $K=8$ capture occasions there are $5^{8}-1=390,624$ possible true
% encounter histories (we ignore the zero history because our model
% conditions on an individual being captured once at least). Of these,
% 325,599 (83.3\%) can occur in the observed data because they contain
% photographs from the left-side only, photographs from the right-side
% only, or at least one simultaneous encounter. The remaining 65,025
% (16.7\%) contain photographs taken from both the left and right sides,
% including via event B, but never simultaneously and are
% unobservable. In practice, we need not consider all 65,025
% unobservable histories. Given a set of observed histories, the
% possible true histories can be constructed by combining each history
% containing only left-side photographs with each history containing
% only right-side photographs as follows: when event L appears in the
% first history and 0 appears in the second enter L in the combined
% history, when 0 appears in the first history and R in the second enter
% R in the combined history, and when L appears in the first and R in
% the second then B is entered into the combined history.

\paragraph{Latent Multinomial Model}
To account for uncertainty in the true encounter histories caused by
multiple marks, we adapt the LMM model of \citet{Link2010}. Suppose
that individuals in the population can have one of $\mtruelink$
possible true histories which produce a total of $\mobslink \leq
\mtruelink$ possible observable histories. The genetic
misidentification model of \citet{Link2010}, for example, allows for
three events on each capture occasion: individuals may be captured and
identified correctly (1), captured and misidentified (2), or not
captured (0). This produces $\mtruelink=3^T$ possible true histories
but only $\mobslink=2^T-1$ observable histories. Following
\citet{Link2010}, we define $\bnobs$ to be the $\mobslink$-vector of
observed counts for the observable histories and $\bntrue$ the latent
$\mtruelink$-vector of counts for the possible true histories. The LMM
is based on two assumptions about these vectors. First, it assumes
that each element of $\bnobs$ is a known linear combination of the
elements of $\bm x$. That is, there is a known $\mtruelink \times
\mobslink$ matrix $\bm A$ such that $\bnobs=\bA\bntrue$. This limits
the possible values of $\bm x$ given the observed value of $\bnobs$,
so we refer to it as the latent vector constraint. Second, the LMM
assumes that $\bntrue$ follows a multinomial distribution
\[
\bntrue \sim \mbox{Multinomial}(N,\bm \pi(\btheta))
\]
with $\ntot=\sum_{k=1}^K \bntrue_k$ representing either the population
or sample size (depending on whether the model conditions on first
capture) and $\bm \pi(\btheta)$, the cell probabilities dependent on
parameter $\btheta$.

% \subsection{Model of True Encounter Histories}

The specific model of $\bntrue$ we have fit is an extension of the
Link-Barker-Jolly-Seber (LBJS) model from \citet{Link2005} modified to
allow for multiple marks. We are primarily interested in the arrival
and departure times of the sharks at NMP and so we condition on
individuals being captured at least one time and ignore the zero
history. In this case, $N$ is the total number of individuals captured
during the study. Note that unlike standard mark-recapture experiments
the true value of $\ntot$ cannot be observed.

The key assumptions of our model are that all emigration from NMP is
permanent, that the probability of remaining at NMP from one occasion
to the next does not depend on how long an individual has been present
(or any other factors), that encounters are independent between
individuals and over time, that there are no losses on capture, and
that the conditional probabilities of the events $L$, $R$, $S$, and
$B$ are constant. Under these conditions, the cell probability
assigned to history $\bm \omega$ is:
\[
\pi_{\bm \omega}(\btheta)= \xi(a|\bm \gamma,\bm \phi,\bm p) \cdot \rho_{\omega_{a}}
\prod_{t=a+1}^{b} \left[\phi_{t-1}(p_t\rho_{\omega_{t}})^{I(\omega_{t}
    \neq 0)} (1-p_t)^{I(\omega_{t}=0)}\right] \cdot \chi(b|\bm \phi,
\bm p)
\]
where $a=\min\{t:\omega_{t}>0\}$ and $b=\max\{t:\omega_{t}>0\}$ denote
the occasions of the first and last captures, and $I(\cdot)$ is the
indicator function. The model is parameterized in terms of:
\begin{enumerate}[1)]
\item {\bf Recruitment rates}: the number of individuals that enter
  the population between occasions $t$ and $t+1$ per individual
  present on occasion $t$ ($\gamma_t$), $t=1,\ldots,T-1$,
\item {\bf Survival probabilities}: the probability that an individual
  present on occasion $t$ is also present on occasion $t+1$
  ($\phi_t$), $t=1,\ldots,T-1$,
\item {\bf Capture probabilities}: the probability that an individual
  present on occasion $t$ is encountered once or more ($p_t$),
  $t=1,\ldots,T$, and
\item {\bf Event probabilities}: the conditional probability of event
  $E$ given that an individual is encountered ($\rho_E$), $E \in
  \{L,R,S,B\}$.
\end{enumerate}
The derived parameter $\xi(a|\bm \gamma,\bm \phi,\bm p)$ models the
probability that an individual is first captured on occasion $a$ given
that it is captured at least one time, and $\chi(b|\bm \phi, \bm p)$
models the probability that an individual released on occasion $b$ is
not recaptured. Expressions for these parameters are provided in
Appendix \ref{app:derived}.  Prior distributions for the model
parameters were chosen to be non-informative whenever possible and are
described in Appendix \ref{app:priors}.

% The survival and capture probabilities were assigned prior
% distributions
% \begin{align*}
% \logit(\phi_t) &\sim N(\mu_\phi,\sigma_\phi^2), \quad t=1,\ldots,T-1\\
% \logit(p_t) &\sim N(\mu_p,\sigma_p^2), \quad t=1,\ldots,T
% \end{align*}
% with
% \[
% \mu_\phi,\mu_p \sim N(0,2) \mbox{ and } 
% \sigma_\phi,\sigma_p \sim HT(3,.9).
% \]
% Here $HT(\nu,\sigma)$ denotes the half $t$-distribution with $\nu$
% degrees of freedom and scale parameter $\sigma$.%  The hyperparameters
% % were chosen because they provide distributions that are close to
% % uniform on $(0,1)$ for both the hierarchical means, $\mu_\phi$ and
% % $\mu_p$, and the base demographic parameters, $\phi_t$ and $p_t$. 
% The recruitment probabilities were assigned prior distributions
% \[
% \log(\gamma_t) \sim N(\mu_\gamma,\sigma_\gamma^2), \quad t=1,\ldots,T-1
% \]
% with
% \[
% \mu_\gamma \sim N(0,.25) \mbox{ and } \sigma_\gamma \sim HT(3,.9).
% \]
% The conditional event probabilities were assigned the prior
% \[
% (\rho_l,\rho_r,\rho_s,\rho_b) \sim \mbox{Dirichlet}(\T{(1,1,1,1)}).
% \]
% Finally, the number of unique individuals encountered was assigned a
% discrete uniform prior on $\{0,\ldots,M\}$ for some $M$ guaranteed to
% be bigger than $\ntot$. In modeling the whale sharks we condition on
% the captured individuals and $N$ is bounded by the number of histories
% observed. All prior distributions were assumed independent.

\subsection{Inference}
\label{sec:methods-inference}

As \citet{Link2010} explains, maximum likelihood (ML) methods are hard
to implement for the LMM. Although the likelihood function can be
written down easily, it is difficult to compute. The distribution of
$\bnobs$ given $\ntot$ and $\btheta$ is a mixture of multinomial
distributions, and its density is easily formulated by summing over
all possible values of $\bm x$ that satisfy the latent vector
constraint. Explicitly:
\[
L(\btheta,N|\bnobs) = \sum_{\{\bm x:\bA \bm x=\bnobs}\} f(\bm x|N,\bm
\theta).
\]
However, there may be many values of $\bm x$ that satisfy these
constraints (even for fixed $\ntot$), and there is no simple way to
identify them all. This makes it difficult to compute the sum directly
and to apply ML inference. Instead, \citet{Link2010} applies Bayesian
inference treating $\bntrue$ as missing data and working with the
joint posterior distribution of $\bntrue$, $\ntot$, and $\btheta$
given $\bnobs$:
\begin{equation}
  \label{eq:4}
  \pi(\bntrue,\ntot,\btheta|\bnobs) \propto
  I(\bnobs=\bA\bntrue)f(\bntrue|\ntot,\btheta)\pi(\ntot,\btheta).
\end{equation}
Inference is then obtained by sampling from this distribution via
MCMC.

The MCMC algorithm that \citet{Link2010} presents is a variant of the
Metropolis-within-Gibb's algorithm which alternately updates the
values of $\btheta$ and $\bntrue$ (note that $\ntot$ is fully defined
by $\bntrue$ and is treated as a derived parameter in the missing data
approach). Updating the value of $\btheta$ given $\bntrue$ is
equivalent to a single MCMC iteration for the parameters of the
underlying mark-recapture model and can be performed with standard
methods. However, it is challenging to update $\bntrue$ given
$\btheta$ in an efficient way. If proposals are generated by making
simple changes to $\bntrue$, e.g. adding or subtracting from randomly
selected elements, then they are unlikely to satisfy the latent vector
constraint and will almost always be rejected. To avoid this problem,
\citet{Link2010} suggests an algorithm that uses vectors from the null
space of $\bA$ to generate proposals for $\bntrue$ that always satisfy
the latent vector constraint. Suppose that $\bm b_1,\ldots,\bm b_R$
form a basis of $\nullA$. Given the current values of $\btheta$,
$\bntrue$, and $\ntot$, call them $\bthetacurr$, $\bntruecurr$ and
$\ntotcurr$, the algorithm updates $\bntrue$ and $\ntot$ by repeating
the following two substeps for each $r=1,\ldots,R$:
\begin{enumerate}[1)]
\item Generate proposals $\bntrueprop$ and $\ntotprop$ by:
  \begin{enumerate}[i)]
  \item sampling $c_r$ from the discrete uniform distribution on
    $-D_r,\ldots,-1,1,D_r$,
  \item setting $\bntrueprop=\bntruecurr + c_r \bm b_r$, and
  \item defining  $\ntotprop=\sum_{r=1}^{\mtruelink} \ntrueprop_r$.
  \end{enumerate}
  \item Compute the Metropolis-Hastings ratio:
    \[
    \alpha(\bntruecurr,\ntotcurr;\bntrueprop,\ntotprop)= \min \left\{1,
      \frac{f(\bntrueprop|\ntotprop,\bthetacurr)\pi(\ntotprop,\bthetacurr)}
      {f(\bntruecurr|\ntotcurr,\bthetacurr))\pi(\ntotcurr,\bthetacurr)}
    \right\}
    \]
    and accept the proposals with probability
    $\alpha(\bntruecurr,\bntrueprop)$.
\end{enumerate}
The key to this algorithm is that $\bA\bm b_r=0$ for each
$r=1,\ldots,R$ so that $\bA \bntrueprop=\bA\bntruecurr+c_r\bA\bm
b_r=\bnobs$. This means that $\bntrueprop$ always satisfies the latent
vector constraint (provided that $\bntruecurr$ also satisfies the
constraint). 
% Moreover, it is theoretically possible to move between
% any two vectors that satisfy the latent vector constraints by adding
% multiples of the vectors $\bm b_1,\ldots,\bm b_R$, so that the chain
% will cover the entire parameter space. 
The values $D_r\in Z$ are
tuning parameters that need to be chosen interactively or before
starting the chain.

Although this algorithm solves the problem of generating valid
proposals for $\bntrue$ and $\ntot$, the computational cost grows
exponentially with $T$. The dimension of $\nullA$ in the genetic
misidentification problem considered by \citet{Link2010} is
$R=3^T-(2^T-1)$. Each update of $\bntrue$ requires 212 substeps if
$T=5$, 58,026 substeps if $T=10$, and $3.5\times10^9$ substeps if
$T=20$.

The amount of computation grows even faster for the problem of
multiple marks. Our model allows for $K=5^T-1$ possible true histories
and $L=(5^T-1)-(4^T-1) + 2(2^T-1)$ observable histories; the dimension
of $\nullA$ is $r=(4^T-1) - 2(2^T-1)$. When $T=8$, there are 390,624
possible true histories of which 325,599 are observable. The MCMC
algorithm of \citet{Link2010} would require 65,025 substeps for each
update of $\bntrue$.

To show how the algorithm can be simplified we consider a toy
example. Suppose that $T=8$ and that only the six histories shown in
the top of Table \ref{tab:example-histories} are observed. These
include two left-only, two right-only, and two simultaneous
histories. Although there are 390,624 possible true histories with
$T=8$ entries, the vast majority of these are not compatible with the
observed histories. In this example, only ten true histories are
compatible with the observed data. These include the six observed
histories plus the four extra histories formed by combining each
left-only and each right-only history, shown in the bottom of Table
\ref{tab:example-histories}. Any other true history would have
produced an observed history not seen in the data.

\begin{table}
  \centering
  \caption{Example of possible observed and true capture
    histories. Suppose that the data comprises the six observed histories given
    in the top of the table. The possible true histories that may have
    generated this data include these six plus the four additional
    histories in the bottom of the table.} 
  \begin{tabular}{lcc}
    & $k$ & History \\\hline\hline
    Observed 
    & 1 & 00L0L000 \\
    & 2 & 0000L000 \\
    & 3 & 00R00000 \\
    & 4 & 000RR000 \\
    & 5 & 00SBR000 \\
    & 6 & S0S00000 \\\hline
    Unobserved 
    & 7 & 00B0L000 \\
    & 8 & 00R0L000 \\
    & 9 & 00LRB000 \\
    & 10 & 000RB000 \\
  \end{tabular}
  \label{tab:example-histories}
\end{table}

Modeling can now be conducted using only the six histories observed
and the ten compatible true histories.  Redefine $\bnobs$ to be the
vector of length 6 containing counts for the observed histories and
$\bntrue$ the vector of length 10 containing counts for the compatible
true histories. The latent vector constraints between $\bnobs$ and
$\bntrue$ are determined by pairing each parent in the observed
histories with its children in the compatible true
histories. Specifically, the number of times a parent is observed must
equal the sum of the counts from all of its children in the compatible
true histories. In the toy example, the first observed history is a
parent of the $1^{st}$, $7^{th}$, and $9^{th}$ compatible true
histories. The corresponding constraint is $f_1=x_1 + x_7 + x_9$. The
remaining constraints are: $f_2=x_2 + x_8 + x_{10}$, $f_3=x_3 + x_7 +
x_8$, $f_4=x_4 + x_9 + x_{10}$, $f_5=x_5$, and $f_6=x_6$.  One
consequence is that $\bntrue$ has only four free elements. New values
of $\bntrue$ can be sampled by updating only $x_7,\ldots,x_{10}$ in
turn and adjusting the remaining counts accordingly. Further, the
values of $x_7,\ldots,x_{10}$ are bounded by the observed counts. In
the example, $0 \leq x_7 \leq \min(f_1,f_3)$, $0 \leq x_8 \leq
\min(f_2,f_3)$, $0 \leq x_9 \leq \min(f_1,f_4)$ and $0 \leq x_{10}
\leq \min(f_2,f_4)$. These bounds can be used to define proposal
distributions that are free of tuning parameters.

Generally, let $\mobs$ denote the number of unique histories observed
and $\mtrue$ the number of compatible true histories. Explicitly,
$\mobs=\mobs_L + \mobs_R + \mobs_S$ and $\mtrue=\mobs +
\mobs_L\mobs_R$ where $\mobs_L$, $\mobs_R$, and $\mobs_S$ denote the
numbers of left-only, right-only, and simultaneous histories
observed. To describe the algorithm we need to know the order of the
counts in $\bm f$ and $\bm x$. We order $\bnobs$ so that the $\mobs_L$
counts of the left-only histories come first, followed by the
$\mobs_R$ counts for the right-only histories, and finally by the
$\mobs_S$ counts for the simultaneous histories. We order $\bntrue$ in
the same way with the counts for the $\mobs_L\mobs_R$ extra,
compatible true histories added at the end. For each of the extra
histories let $\lhistk$ and $\rhistk$ be the indices of its left and
right parents. In the toy example, $l(7)=1$ and $r(7)=3$. The latent
vector constraints are then given by the constraints on the left-only
histories:
\begin{align*}
f_j&=x_j + \sum_{\{k: l(k)=j\}}\ntrue_k,\quad
j=1,\ldots,\mobs_L,\\
\intertext{the constraints on the right-only histories:}
f_j&=x_j + \sum_{\{k: r(k)=j\}}\ntrue_k,\quad
j=\mobs_L+1,\ldots,\mobs_L+\mobs_R,\\
\intertext{and the constraints on the simultaneous histories:}
f_j&=x_j,\quad j=\mobs_L+\mobs_R+1,\ldots,\mobs.
\end{align*}
These equations show that $\bntrue$ is completely defined by the
$\mobs_L\mobs_R$ elements $x_{\mobs+1},\ldots,x_{\mtrue}$ and that
$x_k \leq \min(f_{\lhistk},f_{\rhistk})$ for each
$k=\mobs+1,\ldots,\mtrue$.

Updates to $\bntruecurr$ given $\bthetacurr$ can then be performed
with the following algorithm. For each $k=\mobs+1,\ldots,\mtrue$:
\begin{enumerate}[1)]
\item Generate proposals $\bntrueprop$ and $\ntotprop$ by:
  \begin{enumerate}[i)]
  \item setting $\bntrueprop=\bntruecurr$,
  \item sampling $\ntrueprop_{k}$ from $\{0,\ldots,\min(f_{\lhistk},f_{\rhistk})\}$,
  \item setting
    $\ntrueprop_{\lhistk}=\ntruecurr_{\lhistk}-(\ntrueprop_{k}-\ntruecurr_{k})$
    and $\ntrueprop_{\rhistk}=\ntruecurr_{\rhistk}-(\ntrueprop_{k}-\ntruecurr_{k})$,
  \item and defining $\ntotprop=\sum_{k=1}^{\mtrue} \ntrueprop_k$. 
  \end{enumerate}
\item Reject the proposals immediately if $\ntrueprop_{\lhistk}<0$ or
  $\ntrueprop_{\rhistk}<0$.
\item Otherwise, compute the Metropolis-Hastings ratio:
  \[
  \alpha(\bntruecurr,\bntrueprop)= \min \left\{1,
    \frac{f(\bntrueprop|\ntotprop,\bthetacurr)\pi(\ntotprop,\bthetacurr)}
    {f(\bntruecurr|\ntotcurr,\bthetacurr))\pi(\ntotcurr,\bthetacurr)}
  \right\}
  \]
  and accept $\bntrueprop$ and $\ntotprop$ with probability
  $\alpha(\bntruecurr,\bntrueprop)$
\end{enumerate}
The advantage of this algorithm is that it uses only $\mobs_L\mobs_R$
steps to update $\bntrue$. For the toy example with 6 observed
histories, $\bntrue$ can be updated in 4 steps. For the 2008 ECOCEAN
whale shark data, $\mobs_L=27$ and $\mobs_R=24$ so the new algorithm
requires only 648 substeps to update $\bntrue$. This is much smaller
than the 65,025 substeps required by the algorithm of
\citet{Link2010}. 

We have implemented the MCMC sampling algorithm for fitting the
multiple MARK model directly in \verb?R? and using the JAGS
interpreter for the BUGS language
\citep{Plummer2003,plummer2011,R2012}. An \verb?R? package providing
functions to format the data and to fit these models is available from
the website of the first author at
\url{www.simon.bonners.ca/MultiMark}. In application to the 2008
ECOCEAN whale shark data, we ran three parallel chains with 10,000
burn-in iterations and 50,000 sampling iterations each. Convergence
was monitored with the Gelman-Rubin-Brooks (GRB) diagnostic
\citep{brooks1998} as implemented in the \verb?R? package CODA
\citep{plummer2006}.

% One challenge we faced in assessing convergence was obtaining initial
% values for the chains. Initial values should be diffuse according to
% the posterior, but the parameters for each chain need to be consistent
% so that the chain does not become stuck. To generate initial values
% for each chain we first simulated random values of $\phi_t$ and
% $\gamma_t$, $t=1,\ldots,T-1$, under one of three schemes -- 1)
% $\logit(\phi_t)\sim N(\logit(.9),.1)$ and $\log(\gamma_t)\sim
% N(\log(.1),.1)$, 2) $\logit(\phi_t)\sim N(,.1)$ and
% $\log(\gamma_t)\sim N(\log(.5),.9)$, 3) $\logit(\phi_t)\sim
% N(\logit(.1),.1)$ and $\log(\gamma_t)\sim N(\log(.9),.1)$. We then
% obtained initial values of $p_t$, $t=1,\ldots,T$, by simulating from
% the posterior distribution of the LBJS model considering only the
% left-side photographs and treating the sampled values of $\phi_t$ and
% $\gamma_t$ as fixed. Finally, we generated an initial value for
% $\bntrue$ by sampling from the posterior distribution given the full
% data set but treating all of $\phi_t$, $\gamma_t$, and $p_t$ as fixed.

%%% Local Variables: 
%%% mode: latex
%%% TeX-master: "../multiple_marks_ms_biometrics"
%%% End: 

% Simulation study

\section{Simulation Study}
\label{sec:simulation-study}

\paragraph{Introduction}

To assess the performance of the model presented in the previous
section we conducted simulation studies under a variety of
scenarios. Here we present the results from two simulation scenarios
which illustrate our main results.

In our simulations, we compared the performance of the new model (the
two-sided model) with two alternatives. First, we fit models using
considering only the data from the left-side photographs (the
one-sided model). Capture histories were constructed by combining all
events that include a left-side photograph, namely L, S, and B,
ignoring all right-side photographs. The models we fit to this data
were equivalent to the LBJS model with prior distributions as given in
Appendix \ref{app:priors}. Second, we fit a Bayesian method of
combining inferences from the two sides under the assumption of
independence as in \citep{Wilson1999} (combined inference). To do
this, we fit separate models to the data from the left- and right-side
photographs and averaged the values drawn on each iteration of the
separate MCMC samplers prior to computing summary statistics. For
example, let $\phi_t^{(k,L)}$ and $\phi_t^{(k,R)}$ represent the
values of $\phi_t$ drawn on the $k^{th}$ iterations of the MCMC
samplers run separately for models of the the left- and right-side
data. Let $\widehat{\Var^{(L)}}(\phi_t)$ and
$\widehat{\Var^{(R)}}(\phi_t)$ be the posterior variances estimated
from all iterations. Combined inference for $\phi_t$ was obtained by
computing the inverse variance weighted average of $\phi_t^{(k,L)}$
and $\phi_t^{(k,R)}$
\[
\phi_t^{(k)}=
\frac{(\widehat{\Var^{(R)}}(\phi_t)\phi_t^{(k,L)}+\widehat{\Var^{(L)}}(\phi_t)\phi_t^{(k,R)})}
{\widehat{\Var^{(L)}}(\phi_t) + \widehat{\Var^{(R)}}(\phi_t)}
\]
and then computing summary statistics from the new chain
$\phi_t^{(1)},\phi_t^{(2)},\ldots$. Credible intervals can then be
computed directly from the new chain without relying on normal
approximations. The mean of the values in the new chain is exactly
equal to the inverse-variance weighted average of means from the
separate chains.

We expected that the new model would provide better inference than the
two alternatives. In particular, we expected that credible intervals
from the one-sided models would be wider than the corresponding
intervals from the two-sided model. We also expected that credible
intervals produced by combined inference would be narrower than the
intervals from the two-sided model but would not achieve the nominal
coverage probability.

\paragraph{Simulation 1}

In the first scenario, we generated data under the assumption that all
events were equally likely given capture
($\rho_L=\rho_R=\rho_B=\rho_S=.25$). We set $T=10$ and generated data
by simulating true capture histories sequentially until 200 observed
capture histories were produced (each true history contributing either
0, 1, or 2 histories to the observed data). Demographic parameters
were simulated from the distributions:
\[
  \logit(\phi_t) \sim N(\logit(.80),.30), \quad
  \logit(p_t)  \sim N(\logit(.80),.30), \quad
  \log(\gamma_t)  \sim N(\log(.25),.30).
\]
% \begin{align*}
%   \logit(\phi_t) &\sim N(\logit(.80),.30)\\
%   \logit(p_t) & \sim N(\logit(.80),.30)\\
%   \log(\gamma_t) & \sim N(\log(.25),.30)
% \end{align*}
A total of 100 data sets were simulated and analyzed. The median
number of true histories simulated before 200 observed histories were
obtained was 164 (min=150,max=180), the median number of unique
individuals observed was 138 (min=127,max=148), and the median number
of captures per individual was 2 (min=1,max=10).

Table \ref{tab:sim1} presents statistics comparing the mean-squared
error (MSE) of the posterior means and the mean width and estimated
coverage probability of the 95\% credible intervals obtained from the
alternative models. The MSE of the two-sided model and the
combined-inference were similar for all parameters and smaller than
those of the one-sided model by between 10\% and 25\%. Credible
intervals for both the one-sided and two-sided models achieved the
nominal coverage rate for all parameters, but the credible intervals
for the one-sided model were wider by approximately 10\%. In
comparison, the credible intervals from the combined inference were
narrower than those of the two-sided model by 20\% or more but failed
to achieve the nominal coverage rate.

% \paragraph{Simulation 2}

% In the second scenario, we simulated data under the same model except
% that the capture probabilities were generated from a logit-normal
% distribution with mean $\logit(.5)$. The result is that capture
% probabilities were lowered and the number of observations per
% individual decreased. From the 100 data sets generated, the median
% number of true histories simulated before 200 observed histories were
% obtained was 196 (min=170,max=235), the median number of unique
% individuals observed at least one time was 131 (min=119,max=143), and
% the median number of captures per observed individual was 1
% (min=1,max=9). In this scenario, the MSEs of the posterior mean
% survival probabilities was similar between the two-sided model and the
% combined inference, but the two-sided model produced higher MSEs for
% both the recruitment and population growth rates. Credible intervals
% from the two-sided model were again narrower than those of the
% one-sided model by approximately 10\%, but coverage probabilities from
% both models were above the nominal rate. Finally, although the
% credible intervals from the combined inference were narrower than
% those of the two-sided model by approximately 20\%, the coverage of
% these intervals was also increased and was near or above the nominal
% value. This result was due to the combination of non-informative
% priors and the relatively small amount of information in the data when
% capture probabilities are low; we discuss this point in Section
% \ref{sec:conclusion}

\paragraph{Simulation 3}

In the second scenario, we simulated data from the same model except
that both marks were seen with probability one each time an individual
was captured ($\rho_S=1$). This represents the extreme situation in
which there is complete dependence between the two marks and no
uncertainty in the true capture histories. In this case, the one-sided
and two-sided models produce identical results. The median number of
histories simulated in the 100 data sets before 200 observed histories
were obtained was 215 (min=204,max=227) and the median number of
captures per observed individual was 2 (min=1,max=10).

Point estimates produced by the two models in this scenario were
almost exactly equal and the MSE of the two models was
indistinguishable (see Table \ref{tab:sim1}). However, there were
clear differences in the interval estimates. While the intervals
produced by combined-inference were, on average, 30\% narrower, the
coverage of these intervals was well below the nominal value.

\begin{table}
  \caption{Performance of the estimates from the two
    simulation scenarios. Each column of the table presents the
    MSE of the posterior mean relative to the
    MSE of the posterior mean of the one-sided model,
    and the median width and estimated coverage probability
    of the 95\% credible intervals for the survival probability
    ($\phi$), recruitment rate ($f$), and growth rate ($\lambda$) for
    one of the three models -- one-sided (OS), two-sided (TS), or
    combined-inference (CI). The models are described
    in Section \ref{sec:simulation-study}.}
  \label{tab:sim1}
  \centering
  \begin{tabular}{llrrr@{\hspace{.5in}}rr}
    & &
    \multicolumn{3}{c}{Simulation 1} & 
    \multicolumn{2}{c}{Simulation 2}\\
    & &
    OS&
    TS&
    CI&
    TS&
    CI\\\hline\hline
    \\
$\phi$ 
& MSE & 1.00 & .89 & .87  & 1.00 & 1.00 \\
& Width & .23 & .20 & .16 & .17 & .12 \\
& Cover & .97 & .96 & .90 & .95 & .84\\
\\
$f$ 
& MSE & 1.00 & .88 & .81 &  1.00 & 1.00\\
& Width & .35 & .31 & .24 & .26 & .18\\
& Cover & .97 & .95 & .90 & .95 & .84\\
\\
$\lambda$ 
& MSE & 1.00 & .88 & .82 &  1.00 & 1.00\\
& Width & .41 & .36 & .29 & .31 & .22 \\
& Cover & .98 & .97 & .95 & .97 & .87\\
  \end{tabular}
\end{table}

\section{Results}
\label{sec:application}

The data provided in the ECOCEAN whale shark library contained a total
of 96 observed encounter histories for the 2008 study period. Of
these, 27 histories (28\%) were constructed from left-side photographs
alone, 24 (25\%) were constructed from right-side photographs alone,
and 45 (47\%) contained at least one encounter with photographs taken
from both sides simultaneously. Along with the model presented in
Section \ref{sec:methods}, we computed inferences for $\bp$, $\bmf$,
and $\bphi$ from the alternative models described in Section
\ref{sec:simulation-study}.

Table \ref{tab:ws1} provides posterior summary statistics for the LBJS
parameters model obtained from the two-sided model. Inferences about
all parameters are relatively imprecise because of the relatively
small number of individuals captured and the low capture
probabilities, but the posterior means follow the expected
patterns. Point estimates for the survival probability (the
probability that a whale shark remains at NMP between occasions) are
at or above .90 in the first two periods, below .70 in the last two
periods, and about .80 in between. The posterior mean recruitment rate
is very high in week two, suggesting that most of the sharks entered
during this period, and lower thereafter. This table also provides
summary statistics for the population growth rate, $\lambda_k=\phi_k +
f_k$, $k=1,\ldots,K-1$, computed as a derived parameter. Although the
95\% credible intervals for $\lambda_k$ cover 1.00 for all $k$, the
point estimates are greater than 1.00 for the first two periods, close
to 1.00 in the next three periods, and less than .75 in the last two
periods. This suggests that the aggregation of whale sharks grew
during the first two periods, remained almost steady during the next
three periods, and declined during the last two periods. This supports
the hypothesis that whale sharks aggregate at NMP to feed after the
major coral spawn which occurred between April 9 and 12 in 2008
\citep[][pg.~33]{Chalmers2008}.

\begin{table}
  \caption{Posterior summary statistics for the demographic parameters
    $\phi_k$, $f_k$, $\lambda_k$, and $p_k$ obtained from the two-sided
    model. The columns of the table provide posterior means followed
    with equal-tailed 95\% credible intervals.}
  \centering
  \begin{tabular}{crrrr}
    % Mean & SD & 2.5% & 97.5% & Mean & SD & 2.5% & 97.5% & Mean & SD & 2.5% & 97.5% & Mean & SD & 2.5% & 97.5%\\
 Occ ($k$)
  & \multicolumn{1}{c}{Survival ($\phi_k$)} 
  & \multicolumn{1}{c}{Recruitment ($f_k$)} 
  & \multicolumn{1}{c}{Growth ($\lambda_k$)}
  & \multicolumn{1}{c}{Capture ($p_k$)}
  \\\hline\hline
1 & 0.90(0.67,1.00) & 0.36(0.00,1.93) & 1.26(0.76,2.83) & 0.23(0.08,0.43)\\
2 & 0.92(0.73,1.00) & 2.40(0.08,6.41) & 3.31(1.00,7.33) & 0.19(0.05,0.33)\\
3 & 0.82(0.54,1.00) & 0.17(0.00,0.72) & 0.99(0.64,1.56) & 0.26(0.15,0.43)\\
4 & 0.77(0.45,0.99) & 0.09(0.00,0.36) & 0.85(0.51,1.20) & 0.22(0.13,0.34)\\
5 & 0.82(0.49,1.00) & 0.23(0.00,0.79) & 1.05(0.63,1.65) & 0.22(0.12,0.36)\\
6 & 0.48(0.14,0.96) & 0.06(0.00,0.29) & 0.54(0.17,1.12) & 0.25(0.14,0.42)\\
7 & 0.66(0.16,0.99) & 0.09(0.00,0.42) & 0.75(0.20,1.28) & 0.20(0.06,0.37)\\
8 & \multicolumn{1}{c}{--} & \multicolumn{1}{c}{--} &
\multicolumn{1}{c}{--} & 0.18(0.03,0.34)\\

  \end{tabular}
  \label{tab:ws1}
\end{table}

Table \ref{tab:ws2} provides posterior summary statistics for the
conditional event probabilities. These results show that sharks were
photographed from both sides simultaneously most often
($\hat{\rho_S}=.45(.36,.54)$) and that the probabilities that an
individual was photographed from either the left or right side only
were similar ($\hat{\rho_L}=.29(.20,.38)$ versus
$\hat{\rho_R}=.21(.13,.29)$).

\begin{figure}
  \centering
  \includegraphics[width=\textwidth]{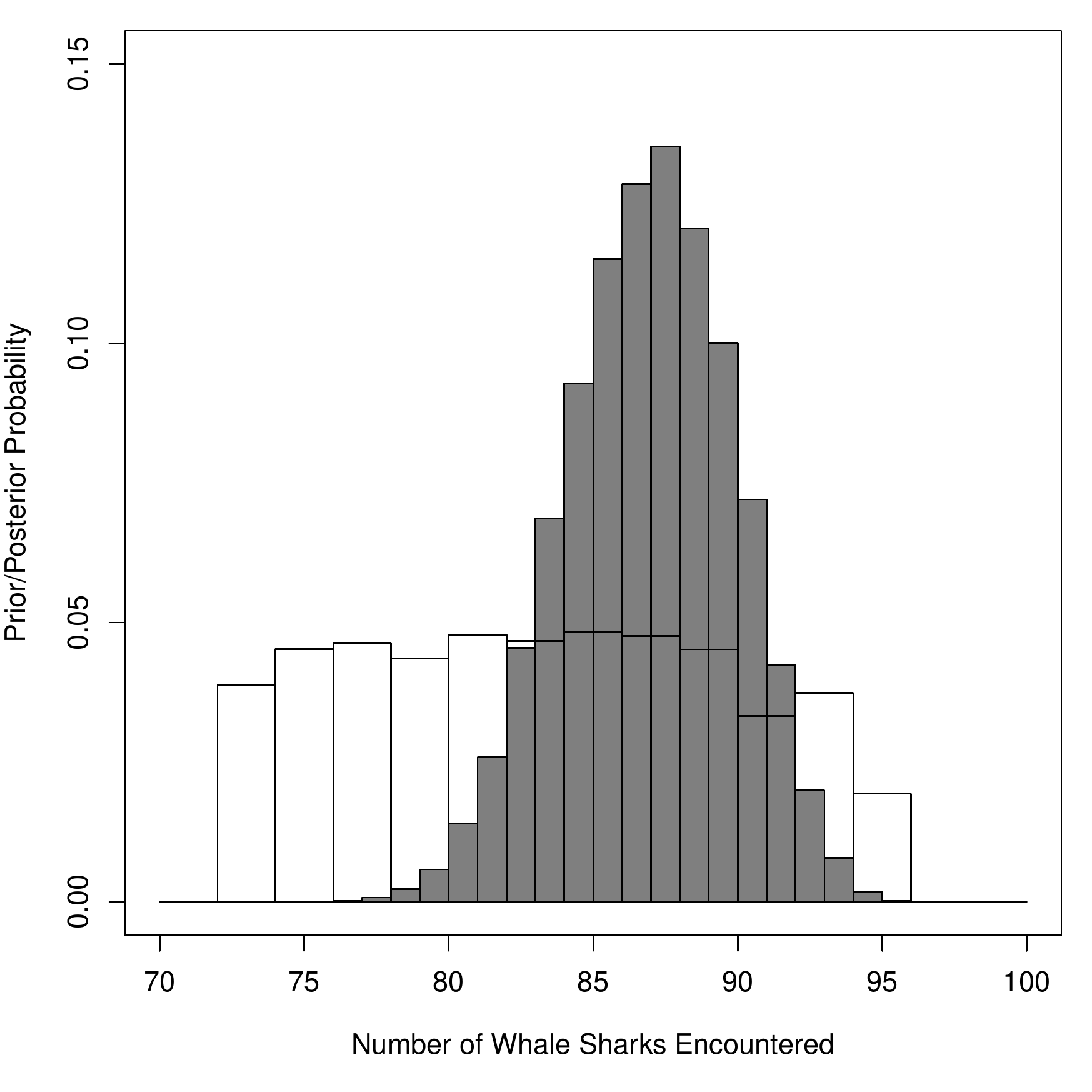}
  \caption{Comparison of the prior and posterior distribution of
    $\ntot$. The prior distribution of $\ntot$, conditional on there
    being 96 observed capture histories and at least 72 unique
    individuals, is shown by the histogram with white bars. The
    posterior distribution of $\ntot$ is shown by the histogram with
    grey bars.}
  \label{fig:ws1}
\end{figure}

The posterior mean of $\ntot$, the number of unique sharks encountered
during the 2008 season, was 88 with 95\% credible interval (82,93).
The full posterior distribution of $\ntot$ is shown in Figure
\ref{fig:ws1} and compared with the prior distribution of $\ntot$
generated by simulating data sets from the prior predictive
distribution conditional on there being 96 observed capture histories
and at least 72 true histories (the minimum number given that 24 of 96
observed histories included right-side photographs alone). Whereas the
prior distribution of $\ntot$ is close to uniform, the posterior
distribution is strongly peaked and concentrates 95\% of its mass
between 82 and 93. We conclude that between 3 (3.1\%) and 14 (14.6\%)
of the sharks encountered during the 2008 season were photographed
from both the left and right sides on separate occasions without ever
being matched.

\begin{table}
  \caption{Posterior summary statistics for the conditional event
    probabilities.}
  \label{tab:ws2}
  \centering
  \begin{tabular}{cc}
    %Mean & SD & 2.5% & 97.5%\\
Event ($j$) & 
\multicolumn{1}{c}{Cond.~Prob.~($\rho_j$)}
\\\hline\hline
1 & 0.29(0.20,0.38)\\
2 & 0.21(0.13,0.29)\\
3 & 0.45(0.36,0.54)\\
4 & 0.06(0.01,0.13)\\

  \end{tabular}
\end{table}

Comparisons of the three chains starting from different initial values
provided no evidence of convergence problems. Traceplots all indicated
that the three chains converged within the burn-in period, GRB
diagnostic values were all less than 1.02, and the estimated MCMC
error was less than 2.6\% of the posterior standard deviation for each
parameter. Based on these results, we are confident that the chains
were long enough to produce reliable summary statistics.

\begin{figure}
  \centering
  %\vspace{6in}
  \includegraphics[width=\textwidth]{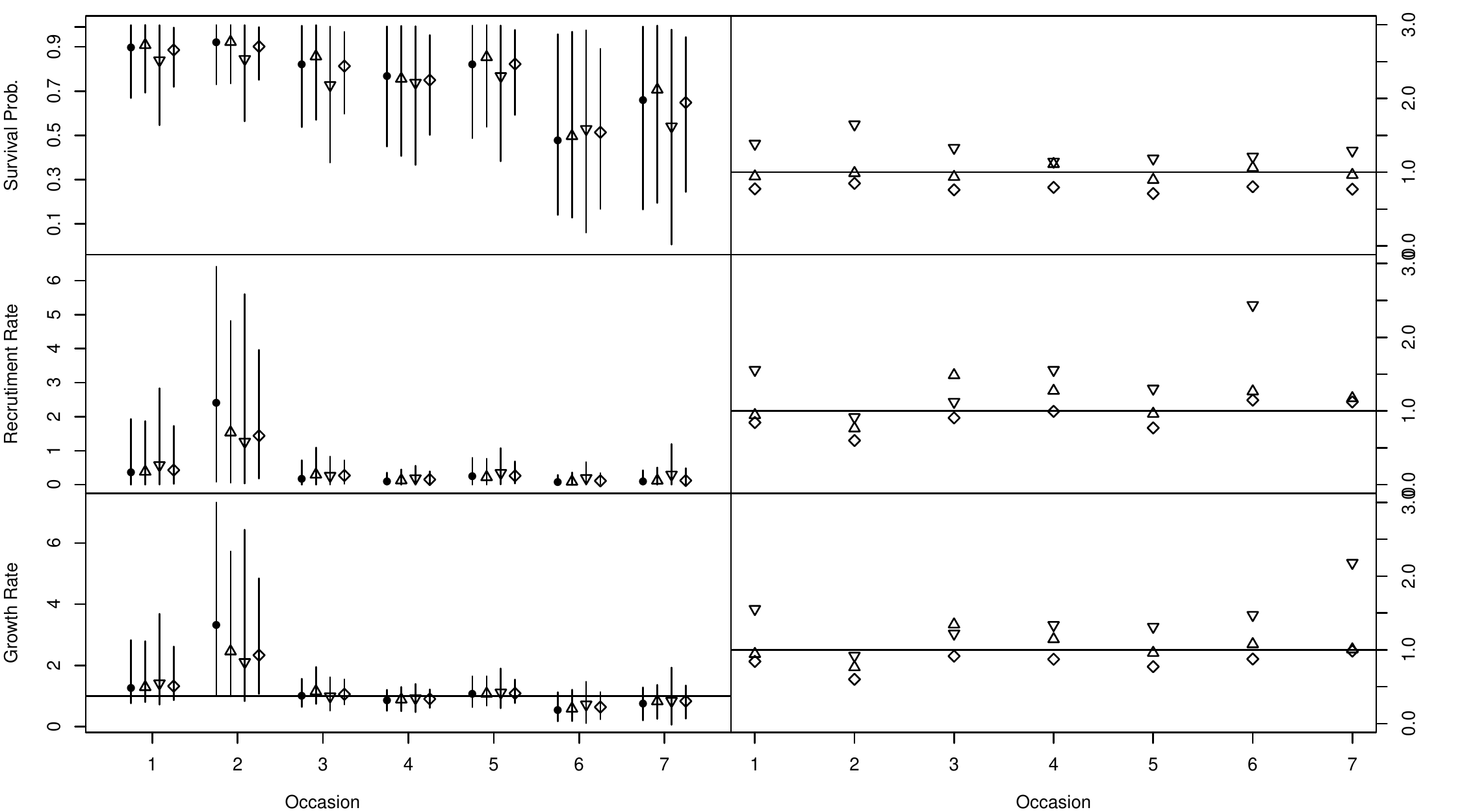}
  \caption{Comparison between the two-sided model and the three
    alternative models. The plots on the left-side of the figure
    compare the posterior means (points) and 95\% credible intervals
    (vertical lines) of the survival probability (top), recruitment
    rate (middle), and population growth rate (bottom) obtained from
    the four models. The plots on the right side of the figure display
    the posterior standard deviations from the three alternative
    models relative to the posterior standard deviation from the
    two-sided model. Results from the two-sided model are represented
    by the circles, from the left-side photographs only by the upward
    pointing triangles, from the right-side photographs only by the
    downward pointing triangles, and from combined inference by the
    diamonds. }
  \label{fig:ws2}
% \end{sidewaysfigure}
\end{figure}

The plots in Figure \ref{fig:ws2} compare inferences for the survival,
recruitment, and growth rates from the four alternative
models. Posterior means from the four models are all very similar and
the 95\% credible intervals for all parameters overlap
considerably. Comparison of the widths of the 95\% credible intervals
from the left- and right-side data alone showed that the two-sided
model provided improved inference for most, but not all,
parameters. On average, the 95\% credible intervals for the
recruitment rates produced by the two-sided model were 93\% and 69\%
as wide as those produced from the left- and right-side data
alone. The 95\% credible intervals for the survival probabilities
produced by the two-sided model were 78\% as wide as those from the
right-side data, on average, but 103\% as wide as those from the
left-side data. This last result seems to be caused by issues with the
upper bound on the survival probabilities as the 95\% credible
intervals for the logit transformed survival probabilities produced
from the two-sided model were, on average, 90\% and 89\% as wide as
those obtained from the left- and right-side data alone. Credible
intervals produced via combined inference were on average 12\% smaller
than those obtained from the two-sided model; however, based on the
results in the previous section, we believe that these intervals would
not achieve the nominal coverage rate and do not reflect the
variability of the parameters correctly.

% \begin{sidewaysfigure}

%%% Local Variables: 
%%% mode: latex
%%% TeX-master: "../multiple_marks_ms_biometrics"
%%% End: 

% Conclusion

\section{Conclusion}
\label{sec:conclusion}

The simulation results presented in Section \ref{sec:simulation-study}
illustrate the main advantages of our model over the previous
approaches to analyzing mark-recapture data with multiple,
non-invasive marks. In general, estimates from our model will be more
precise than estimates based on only one mark. In contrast, the
apparent gain in precision from combining estimators computed
separately for each mark under the assumption of independence is
artificial and credible/confidence intervals computed by these methods
will not achieve the nominal coverage rate. The effect is strongest
when the probability that both marks are seen simultaneously is high
and the separate estimators are highly dependent.

% The differences between the methods are smaller when the capture
% probabilities are low (as in Simulation 2) but this is not surprising
% given that we have selected non-informative prior distributions. The
% posterior distribution is influenced more by the prior as the amount
% of information in the data decreases, and this would not have occurred
% if more informative priors had been selected.

The disadvantage of combining data from multiple marks is that the
model is more complex and computations take longer. A single chain of
60,000 iterations for the 2008 whale shark data implemented in native
\verb?R? code ran in 28.6 minutes on a Linux machine with a clock
speed of 2.8~GHz. In comparison, a chain of the same length for the
one-sided data finished in 6.2 minutes. Our algorithm is less complex
than that of \citet{Link2010}, but the amount of computation is still
proportional to the square of the number of observed histories and the
chains may take too long to run for some large data sets. We are
exploring possible solutions including developing more efficient
methods of computation and approximating the posterior distribution.

% Note that the ratio
% between the median of the credible intervals from the one-sided model
% and the combined inference is close to 1.4 for all parameters in both
% simulation scenarios. This is very close to the $\sqrt{2}$ reduction
% in standard deviation that occurs when two identically distributed and
% perfectly correlated random variables are averaged. We conjecture that
% the posterior standard deviations computed from combined inference
% (assuming independence) will decrease relative to the square root of
% the number of marks used. Of course, the gain is artificial and the
% coverage of credible/confidence intervals will continue to
% decrease. In comparison, our model will account for the dependence
% between the marks and provide more precise inference than the
% one-sided model without sacrificing coverage.

Although we have described our model for two marks, it can easily be
extended for data with any number of marks. We expect that including
more marks will strengthen differences between our model, the
one-sided model, and combined inference seen in the simulation
study. The model can also be adapted easily to estimate the size of an
open population. Following \citet{Link2010}, one can include the null
encounter history (vector of 0s) in the set of possible true
histories. Then $\bntrue$ would have length $\mtrue=\mobs +
\mobs_L\mobs_R + 1$ and $\ntot=\sum_{k=1}^{\mtrue} \ntrue_k$ would
denote the total population size. Because the observed histories do
not restrict the number of individuals never encountered the
constraints on $\bntrue$ would not change. The only differences are
that the MCMC algorithm presented in Section
\ref{sec:methods-inference} would require one more substep to update
the number of individuals never encountered and that the prior bound
on $\ntot$ must be increased to allow for the unobserved individuals.

Non-invasive marks are especially useful for mark-recapture studies
that rely on public data collection because they can often be observed
without special equipment or physical interaction. So called citizen
science projects involving ``public participation in organized
research efforts'' \citep[][pg.~1]{dickinson2012b} play an important
role in ecological monitoring. Large teams of volunteer researchers
can cover large geographical areas and quickly collect large data
sets. As examples of successful, large scale, citizen science projects
in the United States, \citet{dickinson2012b} highlights the US
Geological Survey's North American Breeding Bird Survey (BBS), the
National Audubon Society's Christmas Bird Count, and projects of The
Cornell Lab of Ornithology at Cornell University. The authors estimate
that ``200,000 people participate in [their] suite of bird monitoring
projects each year'' \citep[][pg.~10]{dickinson2012b}.

One concern with many citizen science projects is the reliability of
the data. Some general issues concerning the accuracy and analysis of
data from citizen science projects are discussed by
\citet{cooper2012}. Though the ECOCEAN library does rely on reports
from untrained observers, it differs from similar projects in that
citizens provide no more than the raw data.  Most importantly, the
contributors do not identify the sharks they photograph. Instead, the
submit their photographs to the library and matches suggested by the
paired computer algorithms are all confirmed by trained researchers
(see Section \ref{sec:data}). Hence, the data does not depend on the
ability of tourists or tour operators to identify spot patterns and
matches can be reconfirmed at any time. Even the reported times that a
photograph was taken can be confirmed from the digital timestamp. For
these reasons, we are confident that errors in the data set are
minimized and that the results provided in Section
\ref{sec:application} accurately reflect the arrival and departure of
sharks from NMP in 2008.

Although we are confident in our results, some of the assumptions of
our model given in Section \ref{sec:methods} may oversimplify the
population's dynamics. Sharks may move temporarily to other areas of
the reef and factors like age, sex, or fitness might affect the length
of time that a shark remains at NMP. The objective of this research
was to develop and illustrate a general method for modeling data with
multiple marks, and we intend to explore more complicated models of
the ECOCEAN data in further work. Changes in survival, fecundity, and
capture over time or among individuals might be accounted for with
covariates or random effects, and temporary emigration might be
modeled with Pollock's robust design \citep{Pollock1982}. We also
intend to model data from multiple years in order to assess changes in
the population over time.

% One aspect of the whale shark study that may not occur in other
% studies is that individuals may be encountered multiple times on each
% occasion. Individuals could only be seen once if the occasions were
% truly instantaneous, and in this case both marks could only be
% observed in the same period if they were seen simultaneously. Event
% $B$ could never occur. Our model could be fit to such data simply by
% specifying, {\it a priori}, that $\rho_B=0$. Other researchers have
% also begun to develop similar methods specific to such data
% (McClintock and Conn, pers.~comm.).

%%% Local Variables: 
%%% mode: latex
%%% TeX-master: "../multiple_marks_ms_biometrics"
%%% End: 

% Acknowledgments

\section*{Acknowledgments}
\label{sec:acknowledgments}

We thank Laura Cowen for providing comments on a previous draft of the
manuscript. Matt Schofield also commented on drafts and provide
valuable discussions during the development of our model. Support for
this work was provided in part by the NSF-Kentucky EPSCOR Grant (NSF
Grant No.~0814194). We are aware that similar methods for analyzing
data with multiple marks are being developed by Brett McClintock and
Paul Conn and by Rachel Fewster, and recommend that our work be cited
together.

%%% Local Variables: 
%%% mode: latex
%%% TeX-master: "../multiple_marks_ms_biometrics"
%%% End: 

%Bibliography
\bibliographystyle{biom}
\bibliography{multiple_marks_ms}

\appendix 

\section{}

\subsection{Derived Parameters}
\label{app:derived}
As in \citet{Link2005}:
\[
\xi(a|\bm \gamma,\bm \phi,\bm p)=\frac{\kappa_a}{\sum_{t=1}^T\kappa_t}
\]
where:
\begin{align*}
  \kappa_1&=p_1\\
  \kappa_2&=(\phi_1(1-p_1) + \gamma_1)p_2\\
  \kappa_{t+1}&=p_{t+1}\left(\frac{\kappa_t(1-p_t)\phi_t}{p_t} +
   \gamma_t\prod_{k=1}^{t-1}(\phi_k+\gamma_k)\right),\quad t=2,\ldots,T-1.
\end{align*}
Similarly:
\begin{align*}
  \chi(t|\bm \phi,\bm p) =(1-\phi_t) + \phi_t(1-p_{t+1})\chi(t+1|\bm
  \phi,\bm p), \quad t=1,\ldots,T-1
\end{align*}
with $\chi(T|\bm \phi, \bm p)=1$. 

\subsection{Prior Distributions}
\label{app:priors}
Parameters in the model of the true histories were assigned the
following prior distributions:
\begin{align*}
  \logit(\phi_t) &\sim N(\mu_\phi,\sigma_\phi^2), \quad t=1,\ldots,T-1\\
  \logit(p_t) &\sim N(\mu_p,\sigma_p^2), \quad t=1,\ldots,T\\
  \log(\gamma_t) &\sim N(\mu_\gamma,\sigma_\gamma^2), \quad t=1,\ldots,T-1\\
  (\rho_L,\rho_R,\rho_S,\rho_B) &\sim \mbox{Dirichlet}(\T{(1,1,1,1)})\\
  \ntot & \sim U\{0,\ldots,U_{\max}\}
\end{align*}
The value $U_{\max}$ must be bigger than the true value of
$\ntot$. This can be achieved by setting $U_{\max}=\sum_{l=1}^L f_l$
when the model conditions on first capture.

Hyperparameters were assigned the prior distributions:
\begin{align*}
  \mu_\phi,\mu_p &\sim N(0,2)\\
  \mu_\gamma &\sim N(0,.25)\\
  \sigma_\phi,\sigma_p,\sigma_\gamma &\sim HT(3,.9)
\end{align*}
Here $HT(\nu,\sigma)$ represents the half $t$-distribution with $\nu$
degrees of freedom and scale parameter $\sigma$. All prior
distributions were assumed independent.

%%% Local Variables: 
%%% mode: latex
%%% TeX-master: "../multiple_marks_ms_biometrics"
%%% End: 

\label{lastpage}

\end{document}